\documentclass[preprint,showpacs,preprintnumbers,amsmath,amssymb,endfloats]{revtex4}

\usepackage{graphicx}
\usepackage{dcolumn}
\usepackage{bm}

\begin{document}

\title{Self-learning Kinetic Monte-Carlo method: application to Cu(111)}

\author{Oleg Trushin}
\affiliation{Institute of Microelectronics and Informatics,
Academy of Sciences of Russia, Yaroslavl 150007, Russia }
\author{Altaf Karim, Abdelkader Kara, and Talat S. Rahman*}
\affiliation{ Department of Physics, Cardwell Hall, Kansas State University, Manhattan, Kansas 66506}

\begin{abstract}

We present a novel way of performing kinetic Monte Carlo simulations which
 does not require an {\it a priori} list of diffusion processes and their associated energetics and reaction rates.
 Rather, at any time during the simulation, energetics for all
possible (single or multi-atom) processes, within a specific interaction range, are either computed
accurately using a saddle point search procedure, or retrieved from a database in which previously
 encountered processes are stored. This self-learning procedure enhances the speed of the simulations
 along with a substantial gain in reliability because of the inclusion of many-particle processes.
 Accompanying results from the application of the
 method to the case of two-dimensional Cu adatom-cluster diffusion and coalescence on Cu(111)
  with
detailed statistics of involved atomistic processes
 and contributing diffusion coefficients attest to the suitability of the
 method for the purpose.

\end{abstract}


* corresponding author: email rahman@phys.ksu.edu

\maketitle

\section{Introduction}

The past decade has witnessed a surge in research activities which aim at bridging the gap in length
and time scales at which a range of interesting phenomena take place. Some examples of such
activities pertain to studies of epitaxial growth, and nanostructuring of materials.
  The aim in such work is
to utilize information obtained at the microscopic level to predict behavior at macroscopic scales.
 There are thus several key tasks to be undertaken, each of which is a challenge in itself.
The first of these is an accurate determination of the energetics and dynamics of the system at the microscopic
 level. For selected systems this may be achieved through {\it ab initio} electronic structure
calculations \cite{yu96}
  which are becoming increasingly feasible
 for complex systems, even though they remain computationally intensive.
A reasonable alternative, albeit not as reliable or accurate, is the application of one of several genres
 of many body interatomic potentials \cite{erc88}.
  With these interatomic potentials it has been possible to carry out
 computational and theoretical studies of a range of surface phenomena using techniques like
 molecular statics and
 molecular dynamics. Molecular dynamics
 simulations in particular are capable of revealing the essential
 details of microscopic phenomena as they unfold as a function of temperature, pressure and
 other global variables but the application is limited in time and length scales. Since most
 thermally activated atomistic processes occur in the range of picoseconds,
 they are best captured with time steps
 in femtoseconds which limits total simulation time to few microseconds. These times are many orders of
 magnitude smaller than processes happening in the laboratory. For example, epitaxial growth and
surface morphological changes take place in minutes and hours
 and are controlled by atomic processes which are infrequent compared to atomic vibrational times of
 picoseconds. The challenge in
   molecular dynamics simulations is to find reliable ways which capture infrequent processes
 and extend to longer time scales with reasonable computational resources.

An alternative to molecular dynamics simulations for examining surface phenomena is offered by the kinetic Monte-Carlo (KMC)
 technique in which the rates of various eligible atomic processes are provided as input \cite{bor75,gil76,vot86}.
 If this input
 is accurate and complete, KMC simulations are in good position to mimic experiments.
 Since the task  of accumulating a complete set of atomic processes
 is non-trivial, standard KMC simulations are typically performed with a set of most obvious
 simple atom or concerted processes as
 input, and all others either ignored or included in approximate ways (e-g bond counting models)
 or added in an {\it ad hoc} manner to fit experimental data. [NEW]With a reduced set of
 barriers, activation energies become effective values rather than actual values which
 may be compared with those obtained from experimental data but may not reveal the intervening microscopic processes.
 This is obviously problematic.  Furthermore,
  it has been shown that novel multiple atom processes may play important role in providing mass
 transport on surfaces such as Cu(100) \cite{rowshear,clustdif}, and Ir(111) \cite{diff_exp2,diff_exp3}.  Any realistic simulation should have a provision for uncovering such processes and including their energetics in evolution of the system.

 To overcome these limitations of the two most common
 approaches for simulating temperature dependent morphological evolutions
 of surfaces and interfaces, several accelerated schemes have been presented
in recent times \cite{vot00,hen01,hen03}. In a set of studies, Voter {\it et al}, \cite{vot02} have concentrated
 on enhancing the time scales achievable in MD simulations through three different strategies: parallel-replica,
 temperature-accelerated dynamics and hyperdynamics. Fichthorn and
 co-workers, in related work, apply the bond boost method \cite{mir03} to extend
 the time scales in their simulations.
The basis principle in these methods
 is to make the system evolve faster, sampling a larger phase space, either through smartly
 connected parallel processors, or application of a boost so that the system can
 overcome energy barriers with relative ease, or by raising the temperature of the system.
 At the very least, novel and infrequent processes may be revealed through such accelerated schemes. The
 main issue is the assurance of one-to-one correspondence between the temporal evolution of the accelerated and non-accelerated
 systems and whether the approach actually leads to a large speed-up for a particular system of interest.
 The reader is referred to the original papers
 for further details and suitability of the techniques, to specific cases.

 Another promising scheme has focussed on the completeness issue of KMC by allowing
 the system to evolve according to single and multiple atom processes of
 its choice.
 The key to the method is the generation of saddle points in the potential energy surface
 and benefits from the advances that Jonsson and co-workers
 \cite{neb,dimer} have made in procedures for extracting
 diffusion paths and energy barriers using efficient search procedures.
 Once a large (sufficient) number of saddle points have been identified, the expectation is that the system will evolve naturally according to its inherent mechanisms.
 The method we propose here is in principle related to the latter approach, with a very important
 difference. We employ a pattern recognition scheme which allows efficient storage and subsequent
 retrieval of information from a database of diffusion processes, their paths and their activation energy
 barriers. The procedure presented here is thus efficient and reliable. The removal
 of redundancies and repetitions in the calculations of energetics of system dynamics speeds up
  the simulations by several orders of magnitude making it feasible for a range of applications.
 Since the generation of the database and its future usage through recognition patterns is
 akin to the simulation procedure learning from itself, we call the technique proposed here self learning KMC (SLKMC). While
 the proposed technique can be applied to any surface systems, our interest is in the examination of
 atomistic phenomena as related to growth on fcc(111) surfaces. This is a challenging surface since
 the lack of surface corrugation makes the energy landscape relatively flat
 with a number of diffusion processes which are equally competitive. Some
  such atomistic processes may include those with multiatoms which are typically ignored in
  standard KMC techniques.
 In this paper we focus our attention on some characteristics of the proposed technique and its
 application to
 homo-epitaxy on fcc(111) surfaces through considerations to of the diffusion
 and coalescence of two dimensional Cu adatom islands on Cu(111).
 The structure of the paper is as follows.
 In the next section we present some essentials of the
  self-learning KMC  framework. This is followed in
section III with results of the applications of the method to examine morphological evolution of
  two dimensional Cu islands on Cu(111). Section IV  contains our conclusions.

\section{Essentials of Self-Learning Kinetic Monte Carlo Method}

 Although the principle of the proposed technique is generally applicable,
 we need a specific surface geometry to illustrate its details. For reasons
 mentioned above our interest is homoepitaxy
 on fcc(111) surfaces. We provide in this section some details of the model system, together
 with an outline of
 the standard kinetic Monte Carlo method for completeness.
 This is followed by a summary of the pattern recognition  and labeling scheme that we invoke
 to obtain a self-learning KMC methodology.

\subsection{MODEL SYSTEM}

  To mimic fcc(111) surface we consider a 2-layer substrate, with periodic
 boundary conditions in the XY plane (which is parallel to the surface), to uniquely identify
 the fcc and hcp hollow sites on the surface. The
 system of interest (such as an adatom island,  vacancy island, or any other nanostructure
 whose morphological evolution or diffusion is to be determined) is placed on top of the substrate.
 In this initial study only occupancy of fcc sites (i.e. hollow sites with no atom in the layer below)
 on the substrate is allowed.
 [NEW]While there is experimental justification for assuming fcc-site occupancy for Cu adatoms on Cu(111)\cite{gie03}, we are
 aware that on Ir(111) atoms may also occupy hcp sites (hollow sites with an atom in the layer below) \cite{wan90}.  Infact, even for homoepitaxial growth on Cu(111) under certain other experimental conditions hcp-site occupancy has been reported \cite{cam00}. Furthermore, adatoms, dimers, and other smaller clusters may use the hcp site as an intermediate \cite{repp2003} one during its motion.  The method we are proposing can easily be generalized
 to include hcp occupancy. We are also assuming that the diffusion is via hopping and that
 during the simulation the substrate atoms are kept fixed.
 This restriction can be removed in future work. For the moment our interest is in
 the  in-plane (2D) motion of adatoms, vacancies and their clusters on Cu(111).

\subsection{SOME INGREDIENTS OF KINETIC MONTE CARLO}

The goal of kinetic Monte Carlo (KMC) is to mimic real experiments through sophisticated simulations.
 For these simulations to be realistic, one has to implement increasingly complex
 scenarios requiring intensive use of state-of-the art software and hardware. At the heart of a KMC
 simulation of the time evolution of a given system lie the mechanisms that are responsible for
 determining the microscopic processes to be performed at any given time. To illustrate the point,
 consider a system containing N particles at a given time with $N_e$ possible types of processes.
 Let us also associate with each process-type (i), $n_i$ the number of particles in the system that are
 candidates for this process-type, the activation energy barrier
 $\Delta E_i$ and a pre-factor $\nu_i$. The microscopic rate associated with process (i), within Transition State Theory
(TST) \cite{tst}, is then,

\begin{equation}
 r_{i}=\nu_iexp{-\Delta E_i/kT}, \\
\end{equation}

 where k is the Boltzman constant, and T the surface temperature.
 The total rate R of the system is further given by:

\begin{equation}
   R=\sum_{i=1}^{N_e} R_i , \\
\end{equation}

where $R_{i}=n_{i}r_{i}$  is the macroscopic rate associated with process-type i.

In KMC simulations, the acceptance of a chosen process is always set to one. However, the choice of a given
 process is dictated by the rates. First, a process-type is chosen according to its probability
 $ p_{i}=R_{i}/R $,
and then a particle is randomly chosen from the set $n_{i}$  to perform this process.

 [NEW]The essential elements of the KMC method are thus the processes 'i'
 and their activation energy barriers $\Delta E_i$ whose determination requires
 that availability of reliable interatomic interaction which may be obtained from first principles
 or from model potentials.
 In this paper, all activation energies
 are determined using the embedded atom method (EAM).
   This
 is a semi-empirical, many-body interaction potential \cite{eam8693}.
  Although the EAM potentials neglect
 the large gradient in the charge densities near the surface and use
  atomic charge density for solids, for the six fcc metals
 Ag, Au, Cu, Ni, Pd, and Pt, and their alloys, it  has
 done a successful job of reproducing many of the characteristics
 of the bulk and the surface systems\cite{eam8693}.

To get back to the issue of the determination of diffusion processes, their paths and their
 activation energy barriers, we should note that several interesting
 and appealing approaches have been proposed in the past few years. These
 methods include the nudged elastic band (NEB) method \cite{neb},
 the step and slide mothod \cite{mir01}
 eigenvector following \cite{eigen}, and temperature accelerated MD \cite{tempac}.
 Each of these methods has its own computational demand and measure of
 accuracy whose balance dictates the choice of the approach.
 For the studies presented in this paper, we find the simple 'drag' method
 to be adequate, as we shall see.
 This is, of course, a rudimentary method in which the moving entity is dragged
 in very small steps towards probable (aimed) final state. The dragged
 atoms is constrained in the direction towards the aimed position while
 the other two degrees of freedom (perpendicular to this direction) and all degrees
 of freedom of the rest of the atoms in the system are allowed to relax. The other
 atoms are thus free to participate in the move, thereby activating
 many-particle processes (when neighbor adatoms start to follow central leading atom).
 In connection with SLKMC, the central atom is always dragged towards one
of its
 vacant fcc site. A  more general way to map out the potential energy surface
 is to use the grid method which has been successful in finding
 non-trivial diffusion paths and saddle points \cite{karim01}.

\subsection{SELF LEARNING KINETIC MONTE CARLO METHOD}

 As we have already mentioned, the limitation of standard KMC is its
 reliance on an {\it ad hoc} choice of processes and hence lack of completeness. For these
 reasons and also because of experimental observations of complex
 and unforeseen processes, the predictive power of KMC is in question. A rethinking of the way we perform KMC has become a necessity.
 Simulations with an a priori chosen catalogue of processes need to be replaced by a continuous
 identification of possible processes as the environment changes. For these innovations in the
  KMC procedure, the local environment is the key issue and
 its complexities need to be exploited. With this in mind we are proposing a methodology
 in which the base ingredient is the collection of local environments of undercoordinated atoms found
 automatically during the simulations and labeled and stored for subsequent usage in the simulation.
 As a concrete example of our approach  we have choosen the fcc(111)
 surface with a six-fold symmetry. For simplicity, we assume that any process in this system
 will involve a central (undercoordinated) atom and atoms in the next 3-shells as illustrated in Fig. 1.
 The motif in Fig. 1 is to serve as a 'cookie cutter' and is placed on all active atoms in the system
 to define their local environment.
We further assume, without loss of generality that any process may be described in terms of the central
 atom moving to a neighboring vacancy accompanied by the motion of any other atom or atoms
  in the 3 surrounding shells.
  The labeling of the surrounding atoms is done in binary and a base ten number is then
 associated with the first shell configuration. The same procedure is followed for atoms in the second and third shell.
 Hence, for an atom in the system to be active (i.e. central atom for a given process),
 it should have a vacancy in its first shell (or an occupancy number less than 63 for the cookie cutter); as illustrated in Fig.1.b.

  Once the atoms are classified as active and non-active and encrypted within
 the 3-shell scheme, we proceed by
 determining all possible processes associated with every active atom.
 Next the determination of the activation energy and pre-factor
 is performed for all processes. Examples of how  processes
 are labeled and stored in the database are
 given in Fig.2. In this figure, full circles represent occupied sites and open circles vacancy sites.
 Fig. 2a illustrates the "diffusion along a step" process where the central atom labeled 1 moves to the
vacant site 2 along the step formed by atoms numbered 30, 15, 6, 7, 19 and 37 in the cookie cutter. The initial configuration
for this process is recorded in base 10 as (48,3968,261120) in the database and shown with the base
 2 label in the figure. The move in Fig. 2a is recorded
 as atoms 1 going to position 2 (1,2) and the activation
 energy barrier for the process in Fig. 2a is found to be 0.31 eV. Similarly, for the multi-atom process
 illustrated in Fig. 2b, the initial configuration in base  10 is recorded along
with the sequence of motion of atoms involved in the process which in this case is
 1 going to 4, 6 to 1 and 15 to 15, which is recorded as (1,4;6,1;15,5). This multi-atom process
 was found during the coalescence of two islands and will be discussed later. Its activation energy
 barrier of 0.595 eV is also recorded with the label.

The  bottleneck for the simulation is the determination of the activation energy and the prefactor
 for all possible processes. Even when we make the widely-used
 assumption that all the processes have the same pre-factor, the calculation of the activation energy is very
 expensive if one needs accurate values. Note that since the activation energy is in the exponential,
 any small variation in the activation energy results in a substantial change in the relative probabilities
 and hence the outcome of the whole simulation.
 In standard KMC these energy barriers are provided as input. If, however, as we and others \cite{hen01}
 are proposing that these barriers be calculated on-the-fly, the process will be sped-up
 if provisions are made to avoid recalculations. In the method proposed here this is achieved
 through the storage of activation energy barriers tagged to specific atomic processes in the database.
 This is the basis of our KMC in which
  "self-learning" is achieved by the system through the ability to: (1) calculate activation energies on the fly;
  (2) store them in a database; and (3) recognize and retrieve them using the labeling described above.
 Step 1 is not new. It was already proposed by Jonsson {\it et al} \cite{hen01} and Voter and
 Montalenti
  \cite{vot02}. Step 2 and 3 are unique to our approach and help remove redundancies in the calculations.
 At any given time, after all
 the processes have been sorted out, a search for the activation energies in the database is launched.
 If a new process is encountered, the actual calculation is performed and this process with its activation
 energy is added to the database. Once the processes are classified and macroscopic rates are calculated,
 we proceed to perform one Monte Carlo step in which a randomly selected process is executed. The entire simulation
 process is summarized in the flowchart (Fig. 3).
  At later times in the
 simulation, when the system encounters environments for which some of the possible processes have been
 met earlier, a retrieval process of the activation energy from the database substitutes the actual
 calculation. This gives a tremendous gain in the execution times as evident in our application to the diffusion of 2D Cu clusters on Cu(111). With modest computational resources, it was possible to carry out
  the simulation for a number of MC steps, large enough
 to provide good statistics. The exact number of steps may vary from problem to problem.

In the next section, we discuss some key features of the database
 collected during an extended simulation along with
 the results obtained from applying the SLKMC to post-deposition analysis of homoepitaxy on Cu(111).

\section{Application of SLKMC to Morphological Evolution of 2D islands on Cu(111)}

Since the devil is generally in the details, we present below results
 of the application of SLKMC to study Cu cluster diffusion and coalescence on
 Cu(111). After giving some specifics of the model system, we present
 an analysis of the database which includes an evaluation of the accuracy of the calculated energy barriers and other factors affecting the simulation (cpu) time.  We also comment of the presence and importance of multiatom processes.
 This is followed by the results and discussion of the diffusion and coalescence of 2D clusters on Cu(111).

\subsection{Model Systems}
[NEW]
In the first example, i.e. the study of the diffusion of 2D adatom islands of Cu(111) we have chosen four specific sizes (19, 26, 38 and 100 Cu atoms) for which we already have results for comparison with a KMC simulation using a fixed data base of logical processes involving single atom periphery diffusion \cite{mrs05}.
For the second application to the process of cluster coalescence,
our model system consists of 2 adatom islands, one consisting of 78 atoms and the other 498 atoms placed on
top of the 2-layer substrate.

\subsection{Examination of the collected database}

 [NEW] Since it is important to check the reliability of the data in the created database,  we have compared
 in Table 1 the energy
 barriers that we obtained for some typical diffusion processes presented in Fig. 4,
 using both the drag and the NEB methods. We also include in the Table values available in the literature.
 The comparison in the Table attests to the reliability of the drag method as compared to the more
 time-consuming NEB procedure. For example, with the drag method we were able to achieve
 speed-up of atleast an order of magnitude in the cpu time for the calculation of the energy barriers, as compared to one in which we applied the spherical repulsion methode \cite{tru04} to
 obtain the final states for a given initial state followed by application of NEB method for
 the calculation of the activation energies.

 As an illustration of the richness of the database that we collect, we plot in Fig. 5 the energy distribution
 of about 5000 diffusion processes which have been accumulated during a simulation
 containing several hundreds of millions of Monte Carlo steps. Note from Fig. 5 that the distribution
 is very wide covering activation energies as small as few tens of a meV to about 1 eV. [NEW] Unlike studies
 in which activation energy barriers are calculated for perfect and periodic systems \cite{kar95}, energy barriers  cannot be classified into groups.  Note that in the calculations of the energy barriers differences are
 introduced when the effect of next nearest meighbors of the local environment is included in the calculation, as we have done.
 Note also that the accumulation of the database does not proceed uniformly with time,
 as reflected in  in the insert of Fig. 6. The SLKMC starts, in this case, by accumulating about 400
 different processes during the very first MC step,
 after which the database is "quasi-saturated" for a certain period of CPU time.
 This is followed by an other
 phase of accumulation of about 600 processes, and so on.  It is clear from the slope in Fig. 6 that when the simulation
 runs with a "quasi-saturated"  database, the number of KMC steps/CPU time increases dramatically.
  During  a heavy build up
 of the database, the yield is about 80 KMC steps per second
 and can go up to several thousands of of KMC steps per seconds as the database saturates. [NEW]The onset new events in the database after certain duration of simulation does raise the issue of measures that would assure that the data base is complete.  So far we have found the database to saturate after runs of about 100 - 500 million MC steps.  Actually, for the systems under study we have rarely found new processes to set in after 10 million time steps [OLEG, ALTAF, PLEASE CHECK].

One of the most important features of the method, as we have seen, is its ability to treat many-particle processes,
 the so called "concerted atomic motion". The recent version of the code allows inclusion
 of simultaneous displacements for  atoms up to the third shell. from our simulations
 of several types of local environments
 (straight steps with kinks, compact islands, fractal like islands) we found that in some cases many
 particle processes play important role in providing atomic transport \cite{mrs}.
 They are especially important in the case of low coordination systems, like fractal islands.
 In such cases atoms are weakly bound and prefer to perform concerted motion rather than single
 atomic jumps.
 Furthermore, their importance increases with decreasing
 size of the cluster. In fact molecular dynamics simulations simulations of a
  10-atom Cu island on Cu(111) at 700K and 900K, show that the island  moves by concerted
 displacement rather than through single atoms motion \cite{ahlam05}.
 We next move move onto examination of the results for two specific
 applications of SLKMC.

\subsection{Morphological Evolution}

\subsubsection{Diffusion of 2D islands}

As a first application of the SLKMC method, we present results
 for the diffusion of 2D Cu islands on Cu(111) of four sizes: 19, 26, 38 and 100 atoms.
 These simulations were performed using 10 million [IN THE MRS PAPER WE SAY 500 MILLION, WHICH IS IT]
 MC-steps at 300K and 500K.
 During the simulation, the position of the center of mass was recorded at
 each MC step along with the performed process. After 10 million MC steps, the
 islands have moved far enough that their diffusion coefficient may be extracted from
 the mean square displacement of the center of mass. In Fig. 7 we show the trace
 of the position of the center of mass on the (x,y) plane for both 19 and
 38 atoms clusters at 300K. Note the dark spots for both cases indicating
 a stick-slip type motion of the center of mass. The corresponding mean square
 displacement, for these two atoms, as a function of time show a linear behavior
 (within  statistical errors) and is shown in Fig.8. The extracted slope from the mean square
 displacement plot gives the coefficient diffusion. In Table 2, we report
 the diffusion coefficient for the four cluster sizes at 300K
 and 500K. Note that the diffusion coefficient increases exponentially
 with temperature. The decrease of the diffusion coefficient with
 the cluster size follows a power law ($D=N^{-1.57}$ at 300K
 and $D=N^{-1.64}$ at 500K), which is in good agreement with previous results \cite{dif_th3}.
 The virtue of our calculation is that the atomic processes leading to cluster diffusion were picked by
 the system itself during the simulation. The frequencies of the contributing processes vary with cluster
 size and, more importantly, with surface temperature.
 [NEW] Although these results have already appeared in
 a conference proceeding \cite{mrs}, we present in Table 3 the frequencies of the various contributing atomistic processes. Included in the table are also the frequencies of events that were found in a parallel simulation carried out using the standard KMC simulation with a predetermined catalogue of 294 (49x6) processes involving single-atom peripheral diffusion /cite{gho03}.  It is interesting to note that for the later type of simulation the 19, 38, and 100 atom cluster hardly moved at 300 K, while the diffusion coefficient for the 26 atom cluster was found to be 0.48 $\AA^2$/sec, to be compared with 0.17$\AA^2$/sec.  At 500 K, the diffusion coefficients (in units of $\AA^2$/sec) for the 19, 26, 38, and 100 atom clusters are found by this standard KMC simulation to be 1.29x$10^4$, 2.24x$10^4$, 2.22x$10^4$, and 0.32x$10^4$, respectively.  For 19 atom cluster the diffusion coefficients by KMC and SLKMC differ by an order of magnitude: single atom periphery diffusion underestimates the mobility of this perfect hexagon.  For the other sizes the differences are also noticeable.  Some insights for these differences can be obtained from Table 3 in which we have summarized the frequencies of the processes that are executed in the two types of simulations. While the most dominant mechanism is that of a single adatom diffusion along the (100)-microfacetted step edge (A) in both SLKMC and standard KMC, there are differences in the frequencies which the various processes are executed in the two types of simulations, and also in the appearance of several new processes like detachment from the corners engulfed by the A and B-type ((111)-microfacetted) step edges \cite{mrs}.  In Table 3 the activation energy barriers obtained from the drag method for each of the processes are also compared with those from the nudged elastic band (NEB) method.  Further details of the processes listed in Table 3, including their nomenclature, local configuration, etc. can be found at (http://www.phys.ksu.edu/~rahman). Note that even though multiple atom processes are not very frequent, they do occur,  and in some cases they may be the rate limiting step for the diffusion of cluster of a particular size.

\subsubsection{Island coalescence}

As a second example of application of the SLKMC method, we present here
 results of simulation of the coalescence process in which two adatom islands join together to form a larger
 island with an equilibrium shape on Cu(111).  This simulation was performed  at 300K using
 a small island containing 78 atoms and with an arbitrary shape, put close
 to a larger island containing 498 atoms with a circular shape. Successive snapshots of the system during
  the SLKMC
 simulation are shown in Fig.9, for a total number of forty millions KMC steps.
 From this figure, one notes that a neck between the two islands forms during the first
 hundred thousand KMC steps, corresponding to a physical time of 0.25s. After this time, the neck grows
 until the two islands form an elongated single island after about 10 seconds. Finally, The shape
 of the island evolves to a quasi-triangle with mostly (111)-steps (A-type), which is a result of the
 assymetry in the activation energy barriers associated with A and B-type steps (see Table 1).
 In order to get an insight in the mechanisms involved in the neck formation,
 we have analyzed the frequency distribution of key processes during the first
 and second hundred thousand KMC steps. [NEW]Three types of processes appear prominent in the coalescence
 of these two clusters:
 kink detachment on an A-type step (2a in Fig.4), the reverse of 2a (Rev. 2a in Table 2)
 also called kink incorporation,
 and diffusion along A-type step (4a in Figure 4). Listed in Table 4 are the frequencies for these processes.
 We note from Table 4 that during the
formation of the neck, kink detachment and kink incorporation count for about 15\%
 of the performed processes, another 70\% involve diffusion along A-type step and
  other single and multiple atom processes including kink-rounding and two atoms
 diffusion along steps constitutes the remaining 15\%. For the second hundred thousand KMC steps, the simulation is mostly
 dominated by diffusion along the A-type step (about 96\%), with about 4\% counting for various
 mechanisms. The important fact to note here is that kink-detachment and kink-incorporation
 contributions drop to almost zero after the neck has been formed. Detailed analysis
 of similar simulations involving islands of various sizes and shapes are actually
 in the processes of being performed and will be published elsewhere.
 A similar process for our simulations of cluster island coalescence are in qualitative agreement with the
  observations made  by M. Giesen et al \cite{giesen} using scanning tunneling microscopy.

\section{Conclusion}

We have addressed the issue of completeness of KMC simulations by proposing a method
 in which the system finds, calculates, and collects the energetics of all possible diffusion processes that the moving entities are
 capable of performing. What separates our technique from others recently proposed is the provision for
 storing and retrieving the environment dependent activation energy barriers from a data base. Examination of the data base
 shows that the simulation proceeds much faster when the set of processes is "quasi-saturated' and
 that after sampling such regions the system has the ability to trigger
 the participation of new diffusion processes requiring enhanced CPU's for the calculation of new activation
 energy barriers. The system eventually settles down, the number of MC steps needed to do this depends
 on the system and the number of entries already in the data base (about $10^7-10^8$
 steps). With the use of the pattern recognition scheme we are able to identify and calculate
 the frequency of occurance of individual single and multiple atom diffusion processes that
 actually participate in the evolution of a particular entity.
 The microscopic details of the processes involved in surface morphological evolution can thus
 be documented for a system that have the freedom to evolve on their own accord. We
 show this through applicationto the diffusion and coalescence of 2D adatom islands on Cu(111)
 for which the simulation began with an empty data base. Once a substantial
 accumulation has occured, the simulation time speeds up by orders
 of magnitudes and allows the calculation of system dynamics for time scales relevant to those
 phenomena happening in the laboratory. Interestingly, the two simple examples
 that we have presented here show that only a few dozen diffusion processes
 are in the end vital for a diffusion event. The question of course is: which ones? Our approach answers this
 question.
 As we have already alluded to, the
 task of calculating diffusion prefactors is still ahead of us. This is particularly important since
 we find many competing processes to differ only slightly in energy and differences in their vibrational entropy contributions
 to the prefactors can make a difference in the ultimate evolution of the film morphology. Another
 important result from our simulations with the open data base is that dynamical evolution of
 the system with prejudged diffusion processes may yield erroneous results. Also, the
 pattern recognition schemes to be a prudent way to develop data base of diffusion
 processes and their energetics. It does involve a lot of work in the beginning but once the data base
 is compiled, it can be used for any type of simulation of the system. Of course, for realistic
 simulations of thin films we need to incorporate exchange and other processes which involve motion
 in 3D. [NEW] We have already alluded to the importance of inclusion of hops and occupancies of the hcp site.
 Such efforts are currently underway.  Infact, preliminary results have already been obtained for the
 diffusion of small clusters (2-10 atoms) in which the SLKMC code with the both fcc and hcp occupancy.  However we leave this for presentation elsewhere.  We have also confined ourselves so far to homoepitaxy.  Our future work will also include application to heteroepitaxy for which only minor changes need to be introduced in the methodology.

\bigskip

\section{ Aknowledgements }

We thank James Evans, Chandana Ghosh and Ahlam Alrawi for helpful discussions.
This work was supported by NSF-CRDF RU-P1-2600-YA-04, NSF-ERC 0085604 and NSF-ITR 0428826.


Figure Captions

Fig. 1: a) The three-shell indexing around the central atom labelled 1;
 b) Signature of a particular 2D cluster configuration in base 2 and base 10.

Fig. 2:  Sample (a) single atom and (b) multiple atom processes involved in the diffusion
 of 2D clusters presented with their specific labels for our database.

Fig. 3: Flowchart for SLKMC simulation

Fig. 4: Selected single atom processes on the two types of steps, A (100-microfacetted)
 and B (111-microfacetted), on fcc(111) surface. Process 1 is kink-detachement-rounding,
 2 is kink-detachement-along step, 3 is adatom detachment from step, and 4 is adatom diffusion along step.
 The labels 'a' and 'b' refer to step A and B, respectively.

Fig. 5: Distribution (percentage) of activation energies of stored processes
 in the database during a SLKMC simulation.

Fig. 6: Variations in the number of KMC steps per CPU time ( i.e. performance)
 and the build-up of the databse as a function of the number of KMC steps (insert).

Fig. 7: Trace of center of mass of 19 and 38 atoms Cu clusters on Cu(111) at 300K
 as obtained from SLKMC simulations ($10^7$ steps).

Fig. 8: Mean square displacement (MSD) for 19 and 38 atoms Cu  clusters on Cu(111),
  as function of time at 300K.

Fig. 9: Coalescence of a small Cu cluster (78 atoms)
 with a larger one (498 atoms) on Cu(111) at 300K, using
 SLKMC ($10^7$ steps).

\newpage

Table 1: Diffusion energy barriers for selected mechanisms as shown in Fig. 4.

\begin {tabular}{c|c|c|c}
\hline
Process    &  Drag Method (eV)  &  NEB Method (eV) & Ref. \cite{kar95} (eV)  \\
\hline
    1a     &   0.68             &   0.66           &   -                  \\
    2a     &   0.53             &   0.52           &   -                  \\
    3a     &    -               &   0.65           &   -                  \\
    4a     &   0.25             &   0.25           &   -                  \\
    1b     &   0.60             &   0.59           &   0.59               \\
    2b     &   0.58             &   0.56           &   0.54               \\
    3b     &   0.68             &   0.67           &   0.67               \\
    4b     &   0.32             &   0.30           &   0.29               \\
\hline
\end {tabular}

\vskip 0.6in

Table 2: Diffusion coefficient for 2D Cu islands on Cu(111) ($\AA^2$/sec)

\begin {tabular}{c|c|c}
\hline
cluster Size    &           300K   &          500K \\
\hline
   19     & 0.196               &   1.67x$10^5$               \\
   26     & 0.170               &   8.05x$10^4$               \\
   19     & 0.117               &   4.27x$10^4$               \\
   19     & 0.016               &   1.02x$10^4$               \\
\hline
\end {tabular}

\vskip 0.6in
Table 4: Frequency of selected processes during the coalescence of 2 islands

\begin {tabular}{c|c|c|c}
\hline
Process    &  Barrier (eV)  &  Frequency ($0-1x10^5$ steps) &  Frequency ($1-2x10^5$ steps) \\
\hline
   2a     &   0.530             &   7.41           &    0.03                 \\
    Rev. 2a     &   0.220               &   8.43           &   0.04                  \\
    4a     &   0.25             &   69.66           &   95. 88                 \\
    others  &   -         &   14.50   &   4.05  \\
\hline
\end {tabular}


\begin{thebibliography}{99}



\bibitem{yu96} B.D. Yu and M. Scheffler, Phys. Rev. Lett. {\bf 77}, 1095 (1996).

\bibitem{erc88} F. Ercolessi, M. Parrinello, and E. Tosatti, Phil. Mag.
A {\bf 58}, 213 (1988); M.S. Daw, S.M. Foiles,
 and M.I. Baskes, Mater.\ Sci.\ Rep.\ {\bf 9}, 251 (1993).
\bibitem{bor75} A.B. Bortz, M.H. Kalos, and J.L. Lebowitz, J. Comp. Phys. {\bf 17}, 10 (1975).
\bibitem{gil76} D.T. Gillespie, J. Comput. Phys. {\bf 22}, 403 (1976).
\bibitem{vot86} A.F. Voter, Phys. Rev. B {\bf 34}, 6819 (1986).
\bibitem{rowshear} O.S. Trushin, P. Salo, T. Ala-Nissila, PHys. Rev. B {\bf 62}, 1611 (2000).

\bibitem{clustdif} P. Salo, J. Hirvonen, I.T. Koponen, O.S. Trushin,
 J. Heinonen, T. Ala-Nissila ,
 Phys. Rev. B {\bf 64} , 161405 (2001).

\bibitem{diff_exp2} S.C. Wang, U. Kurpick, G. Ehrlich,
PHys. Rev. Lett. {\bf 81} 4923 (1998).

\bibitem{diff_exp3} S.C. Wang, G. Ehrlich,
Phys. Rev. Lett. {\bf 79}  4234 (1997).


\bibitem{vot00}  M.R. Sorensen and A.F. Voter, J. Chem. Phys. {\bf 112}, 9599 (2000).
\bibitem{hen01} G. Henkelman and H. Jonsson, J. Chem. Phys. {\bf 115},
 9657 (2001).

\bibitem{hen03} G. Henkelman and H. Jonsson, Phys. Rev. Lett. {\bf 90},
 116101 (2003).



\bibitem{vot02} A.F. Voter, F. Montalenti, T. C. Germann, Annu. Rev. Mater.Res. {\bf 32}, 321 (2002).



\bibitem{mir03} R.A. Miron and K.A. Fichthorn, J. Chem. Phys. {\bf 119}, 6210 (2003); R.A. Miron and K.A. Fichthorn,
 Phys. Rev. Lett. {\bf 93}, 138201 (2004).

\bibitem{neb} H. J\'onsson, G. Mills and K. W. Jacobsen,
 in {\it Classical and Quantum Dynamics in Condensed Phase Simulations}, ed. by B. J. Berne {\it et al} (World
Scientific, Singapore, 1998).
\bibitem{dimer} G. Henkelman and H. Jonsson, J. Chem. Phys. {\bf 115}, 7010 (1999).
\bibitem{repp2003} J. Repp, G. Meyer, K. H. Rieder, and P. Hyldgaard, Phys. Rev. Lett. {\bf 91}, 206102 (2003).
\bibitem{gie03} M. Giesen and H. Ibach, Surf. Sci. {\bf 529}, 135 (2003).
\bibitem{cam00}J. Camarero, J.D.L. Figuera, J.J.D. Miguel, R. Miranda, J.
Alvarez, S. Ferrer, Surf. Sci. {\bf 459}, 191 (2000).
\bibitem{wan90} S.C. Wang, and G. Ehrlich,Surf. Sci. {\bf 239}, 301 (1990).
\bibitem{tst} S. Glasstone, K.J. Laidler, and H. Eyring, {\it The Theory of Rate Processes}
 (Mc Graw-Hill, New York, 1941); D. A. King, J. Vac. Sci. Technol. {\bf 17}, 241 (1980).
\bibitem{eam8693}S.M. Foiles, M.I. Baskes, and M.S. Daw,\ Phys. Rev. B
 {\bf 33}, 7983 (1986).

\bibitem{mir01} R.A. Miron and K.A. Fichthron, Molecular Simulation {\bf 30}, 273 (2004).

\bibitem{eigen} L. J. Munro and D. J. Wales, Phys. Rev. B {\bf 59}, 3969 (1999), and references therein.
\bibitem{tempac} A.F. Voter, J. Chem. Phys. 106, 4665 (1997).
\bibitem{karim01} T.S. Rahman, A. Kara. A. Karim, and A. Al-Rawi,
 in {\it Collective Diffusion on Surfaces: Correlation
 Effects and Adatom Insteractions}, ed. by M.C. Tringides and Z. Chvoj (Kluwer Academic
 Publishers, Netherlands) 327 (2001).

\bibitem{tru04}O. S. Trushin, P. Salo, T. Ala-Nissila, and S. C. Ying
Phys. Rev. B 69, 033405 (2004).


\bibitem{mrs} T.S. Rahman, A. Kara, A. Karim, and O. Trushin, Mater. Res. Proceedings (2004).
\bibitem{ahlam05} A. Al-Rawi and T.S. Rahman, unpublished.
\bibitem{dif_th3} A. Bogicevic, C. Liu, J. Jacobsen, H. Metiu,
Phys. Rev. B 57, R9459-R9462 (1998).
\bibitem{giesen} M. Giesen
( internal report)
\bibitem{kar95} M. Karimi, T. Tomkowski, G. Vodali, O. Biham,
Phys. Rev. B {\bf 52}, 5364 (1995).
\bibitem{gho03} C. Ghosh, A. Kara, and T. S. Rahman, to be published; C. Ghosh. PhD thesis, Kansas State University, 2003.























\end{thebibliography}
\end{document}